\documentclass[fleqn,usenatbib]{mnras}

\usepackage{newtxtext,newtxmath}

\usepackage[T1]{fontenc}

\DeclareRobustCommand{\VAN}[3]{#2}
\let\VANthebibliography\thebibliography
\def\thebibliography{\DeclareRobustCommand{\VAN}[3]{##3}\VANthebibliography}

\usepackage{graphicx}	
\usepackage{amsmath}	

\title[EBC for high speed optical astronomy]{Neuromorphic cameras for Atmospheric Cherenkov Telescopes and fast optical astronomy: new paradigm, challenges and opportunities}

\author[J. Hoang]{
J. Hoang$^{1,2}$\thanks{E-mail: jokhoang@ucsc.edu (SCIPP)}
\\
$^{1}$Santa Cruz Institute for Particle Physics, Natural Sciences 2 Building, 1156 High Street, Santa Cruz, CA 95064\\
$^{2}$Berkeley SETI Research Center, Radio Astronomy Lab, Campbell Hall, University Drive, Berkeley, CA 94720 - 3411\\
}

\pubyear{2015}

\begin{document}
\label{firstpage}
\pagerange{\pageref{firstpage}--\pageref{lastpage}}
\maketitle

\begin{abstract}
The astronomy community has witnessed an explosive growth in the use of deep-learning techniques based on neural networks since the mid-2010s. The widespread adoption of these nature-inspired technologies has helped astronomers tackle previously insurmountable problems and provided an unprecedented opportunity for new discoveries. However, one of the primary tools of today's optical astronomy is neither natural nor efficient: their photo-sensing devices. Specifically, the modern CCD camera - like that of the cutting-edge Rubin Observatory - requires an internal clock to regularly expose the sensor to light, consumes a large amount of energy and information bandwidth, and has a limited dynamic range. On the contrary, biological eyes lack an internal clock and a shutter, have much higher pixel density but consume significantly less energy and bandwidth, and can adapt to bright and low light conditions. Inspired by the nature of the eyes, M. Mahowald and C. Mead introduced the revolutionary concept of a silicon retina sensor in 1991. Also known as event-based cameras (EBCs), these types of devices operate in a vastly different way compared to conventional CCD-based imaging sensors. EBCs mimic the operating principles of optic nerves and continuously produce a stream of events, with each event generated only when a pixel detects a change in light intensity. EBCs do not have fixed exposure times, have high dynamic range, require low power for operation, and can capture high-speed phenomena. These properties are important requirements for Cherenkov telescopes as well as other high-speed optical astronomy. This work presents the opportunities and challenges of using EBCs in those cases, and proposes a low-cost approach to experimentally assess the feasibility of this innovative technique. 
\end{abstract}




\section{Introduction}

Cherenkov telescopes are a special class of optical telescopes designed to indirectly observe gamma rays from astrophysical sources, and the Imaging Atmospheric Cherenkov Technique/Telescope (IACT) is now an established field. Traditionally, IACTs have relied on the conventional imaging paradigm to capture the nanosecond-duration cascade of Cherenkov photons produced by gamma-ray interactions with the Earth's atmosphere. In order to capture the fleeting and faint physics of an atmospheric airshower, each telescope employs a special type of camera equipped with fast photomultipliers instead of CMOS/CCD photosensors. IACT cameras do not continuously capture snapshots of the sky at regular nanosecond intervals: the data volume produced by doing so would be unmanageable. Instead, through a custom-made hardware triggering scheme, the camera only samples data during a window of $\sim$100ns when a group of pixels (usually three neighboring pixels) meets predetermined criteria, including a specific threshold and pattern. Depending on the camera design, the output is either a movie of the shower development or just the stacked image of the airshower event, as seen by the entire camera. Subsequent data analysis steps involve cleaning the camera image to isolate the airshower image from the background and infer the properties of the primary incoming particle based on the geometry of the cleaned images.

Recently, an emerging type of camera inspired by the neural architecture of the eye has gained significant attention in the field of computer vision. Often referred to as Event-Based Cameras (EBCs) in literature, EBCs are a neural-inspired class of photosensors that operates on the principles of the photoreceptor neurons in biological eyes \citep[]{silicon_retina}. Instead of capturing synchronously all pixel values at a fixed sampling rate like the traditional frame-based camera, EBCs operate asynchronously based on the per-pixel change in brightness level. EBCs generate a stream of event tuples, each consisting of a timestamp (T), pixel coordinates (X,Y), and a +/- P polarity sign indicating brightness changes. EBCs are advantageous where the scenarios are challenging for traditional cameras, such as high speed, object isolation, and high dynamic range; they have found applications in astronomy and space situational awareness such as stars or satellite tracking \citep[]{EBC_star_tracking,EBC_satellite_tracking}. However, the use of EBCs in IACT has not been explored, despite the fact that IACT cameras are designed to specifically look for faint and fast cosmic-ray airshower “events” where the use of EBC makes more logical sense. High speed is a key requirement for any IACT camera: significant efforts have been put into ensuring this requirement through the use of specialized detectors and custom-made readout electronics. Object isolation simplifies the data analysis algorithms used in IACT. A high dynamic range is a valuable addition for moonlight observations. 

In addition, IACTs are still fundamentally optical telescopes; with their large 10-m-class tessellated reflecting surfaces, they are effectively large “photon buckets” and make the instrument suitable for several direct optical observations. However, these observations often require several technical modifications to both hardware and software to decouple the photosensor from the standard shower-triggered regime so that the sensor can fully leverage its fast-timing nature. Constrained by resources, only one sensor per telescope is typically utilized during these non-standard optical observations. Having multiple fast sensor readouts per telescope can therefore offer immense advantages. EBC can serve as the bridge closing the MHz gap between IACT and traditional kHZ optical astronomy. 

The remaining parts of this work will explore the potential use of EBT in IACT and fast optical astronomy, and suggest a low-cost pilot concept for an EBC-based telescope to be used in both gamma-ray as well as optical astronomy.

\section{Advantages of EBCs}\label{sec:High_Temporal_Resolution} 
Due to the different operating principles, EBC offer several potential advantages over standard photomultiplier-based cameras in IACT and CCD-based conventional cameras in optical observatories.

\subsection{Low data volume}\label{ssec:2.1}
In the imaging paradigm, airshower images are usually contained in a fraction of the camera plane. However, the value of every pixel is recorded during an airshower, and the data analyzer must subsequently select only the relevant pixels using cleaning algorithms. On the contrary, EBC streams a sequence of (X, Y, T, P) tuples only when the light level changes, presumably due to the rising or falling flux of photons. Thus, not all pixel values will be recorded and read out during each shower event. The captured images are segmented i.e. only pixels containing shower photons are recorded. This built-in image segmentation also affords us greater spatial resolution.

\begin{figure}
    \includegraphics[width=0.5\textwidth]{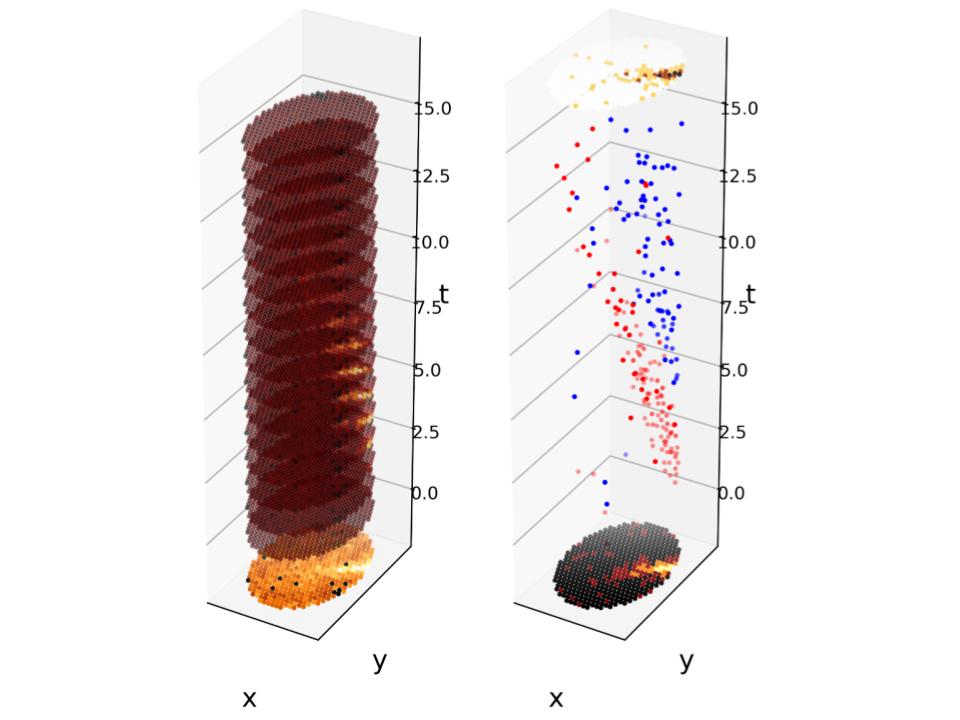}
    \label{fig:Sample}
    \centering
    \caption{Comparison between frame-based (left) and event-based (right) paradigms on the same shower event. See Section \ref{ssec:2.1} for explanations.}
\end{figure}

Fig.\ref{fig:Sample} illustrates how EBCs can reduce their data volume even though they operate at a higher speed. The left plot shows a candidate gamma-ray shower as seen by a VERITAS telescope. It consists of 16 frames separated by 1 ns with each frame having 499 pixels, thereby requiring a (499$\times$18) array to store the shower information. The bottom frame at $z = -2$ is the integrated (stacked) image, which is typically used during the analysis. The right plot shows a simulated response of an EBC operating at a 0.1 ns speed. Red dots denote positive polarity where the difference in \textit{log(charge)} exceeds 0.5. Blue dots are where it falls below 0.5. The bottom frame is the stacked image of positive polarity points, and the top frame is the stacked image of negative polarity points. For this particular shower, there are 125 points with positive polarity and 81 points with negative polarity. Each red/blue point requires a pixel ID, a timestamp, and a $\pm$ polarity value, i.e. a (206$\times$3) array in total. We therefore achieve a reduction of $\sim$ 13 times in data volume for this shower event. This factor can be further increased if only one polarity is used.     

\subsection{High dynamic range}\label{ssec:2.2}
With a high dynamic range of >100 dB, EBCs can capture information from low-light to daylight conditions. Like biological retinas, EBC can self-adapt to very dark as well as very bright stimuli. On the contrary, the dynamic range of a SiPM is typically limited to $\sim$50 dB, and an IACT camera can therefore operate only under dark to moderate moonlight conditions in order to avoid pixel saturation. Low dynamic range limits the number of observation hours of the instrument. Replacing the IACT SiPM-based camera with a high dynamic range EBC will significantly improve the instrument's duty cycle.

\subsection{High Temporal Resolution}\label{ssec:2.3}
With readout rate ranging from 2 MHz to 1200 MHz \citep[]{EBC_survey} EBCs can theoretically monitor the change in brightness in the order of 500 ns to 0.8 ns. The smaller 0.8 ns window falls within the Cherenkov regime, where the night sky background (10 - 100 MHz) no longer dominates. The larger 500 ns window, although too slow for direct gamma-ray observations, is still useful for other fast optical astronomy cases such as searching for optical bursts from FRBs, optical pulsar pulsations, or stellar occultations.

\subsection{Low voltage, low power}\label{ssec:2.4}
Since EBC transmits only changes in brightness, power is only consumed to process changing pixels. At the die level, most EBCs use about 10 mW and require an operating voltage of a few volts. For comparison, the PMT-based MAGIC-II camera typically consumes around 1 kW of power and requires up to 1.2 kV to operate the PMTs \citep[]{MAGICII_peroformance} during operation. Newer SiPM-based IACT cameras reduce these figures drastically, but will not match EBC values due to the presence of ASIC/FPGA electronic boards for triggering and signal digitization.

\subsection{Cost and availability}\label{ssec:2.5}
Due to their complexity, IACT cameras are expensive to produce and are currently exclusively manufactured at research institutes, often involving multi-team collaborations. On the other hand, there are already several commercially available EBCs. While vendors do not typically disclose their product prices to the public, it is estimated to be several thousand US dollars, significantly lower than the full cost (material + manpower) to build an IACT camera.

\section{Challenges and open questions on the implementation of EBC in IACT}\label{sec:Challenges} 
As is the case with new technologies, there are several unknown characteristics that can pose a challenge to the implementation of EBC in IACT.

\subsection{Sensitive estimation}\label{ssec:3.1}
\begin{figure}
    \includegraphics[width=0.5\textwidth]{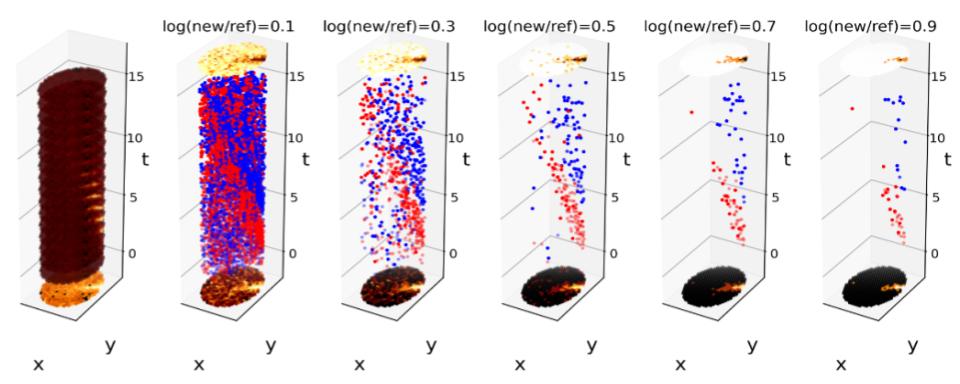}
    \label{fig:threshold}
    \centering
    \caption{Threshold effects on sensitivity and bandwidth. A high threshold will lead to a decrease in sensitivity as pixels will not trigger on a faint event, while a low threshold will improve sensitivity but may saturate the bandwidth.}
\end{figure}

While conventional IACT cameras react to absolute units i.e. photon counts, EBTs are sensitive to differential units i.e. brightness change. Positive (“ON”) and negative (“OFF”) events may be triggered according to different programmable thresholds, denoted as C+ and C-, which set the speed and threshold voltages of the change in the detector \citep[]{EBC_survey}. Changing these thresholds may increase/decrease the camera’s sensitivity. Typical EBCs can set the thresholds between 10\% and 50\% illumination change, although 1\% is possible at the expense of increased noise and bandwidth consumption. The simulation depicted in Fig. \ref{fig:threshold} illustrates the impact of adjusting the threshold on the camera's sensitivity. The optimal threshold for shower-based physics as well as how changing the C+/C- threshold affects the sensitivity is currently unknown. It is also unclear whether it is sufficiently sensitive to a flux change of a few Cherenkov photons from a low-energy shower.

\subsection{Timing resolution}\label{ssec:3.2}
EBCs have readout rates ranging from 2 MHz to 1200 MHz \citep[]{EBC_survey}. However, it is unclear how this rate actually is measured when several pixels are triggered during an airshower, and what the actual bandwidth of the camera is while being used during Cherenkov observation. The timestamps are typically accurate to microsecond resolution, which is about an order of magnitude lower than what is required for conventional IACT standards. It remains unclear whether improvements in hardware or algorithms can overcome this limitation. Additionally, the relative timing consistency between pixels is unclear; this limitation will affect the precision of shower reconstruction.

\subsection{Data interpretation}\label{ssec:3.3}
The data generated by EBC are fundamentally different from the conventional image-based camera. Therefore, totally different algorithms for event reconstruction and analysis are required. Such an algorithm must also take into account the intra-pixel timing uncertainty. Developing and optimizing these algorithms can be a complex and computationally intensive task. In addition, it is unclear whether the sparse data volume contains sufficient information to determine crucial physical properties of the incoming primary particle such as arrival direction and energy. Finally, similar to our biological eyes, EBCs may suffer from optical illusions such as ghosting, afterimages, and grid illusions. Further studies must be conducted to understand how these illusions affect observational data. 

\subsection{Integration}\label{ssec:3.4}
Retrofitting EBC into existing IACT hardwares may require significant modifications and calibration to ensure compatibility with the new technology. In addition, telescope pointing stability must be ensured, as EBC can easily detect minute movements such as vibrations. On the software side, special algorithms must be developed to remove unwanted events, such as optical trails from satellites and meteors \citep[]{HESS_trail, MAGIC_cpix}.

\section{Some use cases of EBC towards a low-cost fast optical astronomy}\label{sec:Targets}

\subsection{Ground-based gamma-ray astronomy}\label{ssec:4.1}
If EBCs do indeed possess the sufficient sensitivity and timing accuracy mentioned in \ref{ssec:3.1} and \ref{ssec:3.2}, they will pave the way toward a novel IACT construction. Additionally, the high dynamic range also enables moon-light observations and extends the duty cycle of the telescopes.

\subsection{Technosignatures detection}\label{ssec:4.2} 
The VERITAS telescopes have also performed a search for technosignatures and reported their observation of laser signals from a weather satellite \citep[]{VERITAS_BL}. The analysis method reported was sub-optimal due to the significant amount of cosmic ray background noise that must be filtered out. Specifically, \cite{VERITAS_BL} only searched for technosignatures in the region around the expected source’s location in the camera plane, but did not extend the analysis to the full camera region. Using EBCs with the direct pixel coordinates where the photon flux has changed will simplify the analysis process, enabling real-time detection for any position within the camera’s field of view. Additionally, the detection of objects in geosynchronous orbits using EBCs has also been reported. LEO objects were also reported even during daytime \cite[]{EBC_satellite_tracking}, demonstrating the advantage of having a high-dynamic range camera. With appropriate filtering techniques, EBCs will allow, for the first time, the first continuous all-sky all-time searches for optical technosignatures.

\subsection{Asteroid occultations}\label{ssec:4.3}
The VERITAS Cherenkov telescopes have performed a direct measurement of stellar angular diameters from occultations using four of its central pixels at data sampling between 300 Hz and 4800 Hz \citep[]{VERITAS_occultation}. EBCs have significantly more pixels and provide an exciting venue to explore an alternative instrumental approach to the method.

\subsection{Other optical targets}\label{ssec:4.4}
The versatile capability of the central PMT of the Chrenkov telescope for optical astronomy has been reported in \cite{MAGIC_cpix}. It can be used to observe a wide range of targets such as optical pulsars, optical counterparts of Fast Radio Bursts and microquasars, and meteor studies. These targets can be studied at higher sensitivity with EBCs by having a significantly larger number of usable pixels. For example, spurious meteor signals were considered irreducible background with the limited use of a single central pixel of the MAGIC camera \citep[]{MAGIC_FRB}. The use of a multipixel EBC will enable better discrimination between ms-duration optical flashes from meteors and genuine optical counterparts from FRBs. 

\section{Proposal for a low-cost IACT-like telescope with EBC}
\begin{figure}
    \includegraphics[width=0.5\textwidth]{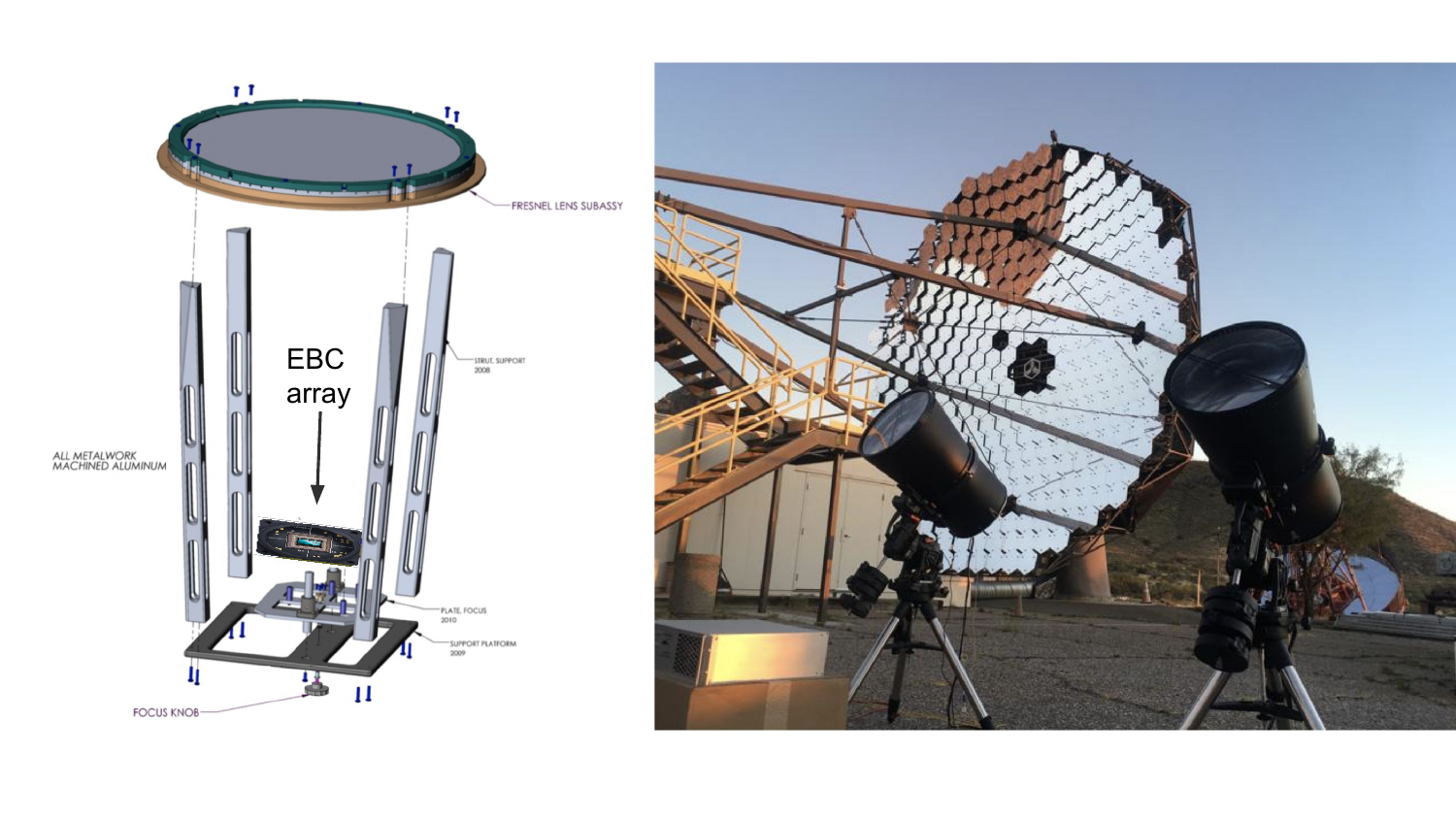}
    \label{fig:Scheme}
    \centering
    \caption{Left: Exploded view of the proposed telescope with an EBC array at the focal point, replacing the SiPM sensor array from the original PANOSETI concept described in \protect\cite{PANOSETI_design}. Right: Two of the PANOSETI telescopes in the foreground and the VERITAS telescope in the background during a joint gamma-ray observation session mentioned in \protect\cite{VERITAS_PANOSETI}.}
\end{figure}

\cite{VERITAS_PANOSETI} proposes a smaller and more affordable telescope concept than traditional IACT, making them suitable instruments to deploy as an array or in fly-eye configuration. Here we propose a modification to the so-called PANOSETI design \citep[]{PANOSETI_design} by replacing the SiPM array with an EBC array, shown in Fig. \ref{fig:Scheme}. Briefly, the optomechanical design of the telescope consists of a Fresnel lens, a photosensor array placed at the focal point of the lens, an adjustable moving stage to fine-tune the position of the array, a baffle to shield the sensor from stray light, and a commercial-grade steerable mount. Instead of using an array of SiPMs with customized electronic boards, an off-the-shelf EBC will be used. The EBC is battery-powered and does not need to be connected to an external power supply. This instrument can be used to study targets mentioned in Section \ref{sec:Targets} of this work.

\section{Conclusions}
Inspired by biological processes, neuromorphic cameras offer a novel and efficient way of dynamically sensing the physical world. The new paradigm eliminates many inherent limitations of the traditional frame-based camera system, offering the potential for high-speed astronomical observations. This work introduces the rationales for using EBCs in fast optical astronomy, including ground-based gamma-ray astronomy. It proposes a low-cost EBC telescope concept and observation targets to assess the potential and limit of this new technology. If proven, neuromorphic eyes have the potential to revolutionize the community similar to the use of deep-learning techniques based on neural networks in astronomy. After all, it is quite true that Nature is the best teacher and the greatest source of wisdom.

\section*{Acknowledgements}
The author would like to thank the Breakthrough Prize Foundation for their support. Special thanks to C. Cheung for the early inspiration. This work is funded by the Breakthrough Listen Initiative.



\bibliographystyle{mnras}



\bsp	
\label{lastpage}
\end{document}